\documentclass{iopart}
\usepackage{graphicx}
\usepackage{amssymb}
\begin{document}
\title{Controlling the order of wedge filling transitions: the role of line 
tension}
\author{J. M. Romero-Enrique}
\address
{Departamento de F\'{\i}sica At\'omica, Molecular y
Nuclear, Area de F\'{\i}sica Te\'orica, Universidad de Sevilla,
Apartado de Correos 1065, 41080 Sevilla, Spain}
\author{A. O. Parry}
\address{
Department of Mathematics, Imperial College 180 Queen's Gate,
London SW7 2BZ, United Kingdom}
\begin{abstract}
We study filling phenomena in 3D wedge geometries paying particular attention
to the role played by a line tension associated with the wedge bottom. 
Our study is based on transfer matrix analysis of an effective 
one dimensional model of 3D filling which accounts for  
the breather-mode excitations of the interfacial height. 
The transition may be first-order or continuous (critical) depending 
on the strength of the line tension associated with the wedge bottom. 
Exact results are reported for the interfacial properties near filling 
with both short-ranged (contact) forces and also van der Waals interactions. 
For sufficiently short-ranged forces we show the lines of
critical and first-order filling meet at a tricritical point. This
contrasts with the case of dispersion forces for which the lines meet at 
a critical end-point. Our transfer matrix analysis is compared with generalized
random-walk arguments based on a necklace model and is shown to be
a thermodynamically consistent description of fluctuation effects at filling. 
Connections with the predictions of conformal invariance for droplet shapes 
in wedges is also made.
\end{abstract}
\maketitle

\section{Introduction} 

Fluid adsorption on micropatterned and sculpted solid substrates 
exhibit novel phase transitions compared to wetting behaviour at planar, 
homogeneous walls \cite{Gau,Rascon,Bruschi}. The simple 3D wedge geometry 
has been extensively studied in the past decade theoretically 
\cite{Rejmer,Parry,Parry4,Parry2,Bednorz,Henderson,Henderson2,
Henderson3,Rascon2,JM1,JM2}, experimentally 
\cite{Bruschi0,Bruschi,Bruschi2,Bruschi3} and 
by computer simulation \cite{Milchev,Milchev2,Binder}. 
Thermodynamic arguments \cite{Concus,Pomeau,Hauge} show that the wedge is
completely filled with liquid provided the contact angle $\theta$
is less than the tilt angle $\alpha$. These studies show that
the conditions for continuous wedge filling transition are less 
restrictive than for critical wetting at planar walls \cite{Parry,Parry4}.
Close to critical filling, the substrate geometry enhances interfacial
fluctuations, which become highly anisotropic. We refer to these as 
breather modes excitations \cite{Parry,Parry4}. 
However, most of these studies neglect the 
presence of a line tension associated with the wedge bottom. Previous 
studies by the authors \cite{JM1,JM2} for short-ranged binding potentials 
show that the line tension may play an important role in filling phenomena
and may drive the transition first-order if it exceeds a threshold value. 
We extend our analysis to arbitrary binding potentials, in 
particular to van der Waals dispersive interactions. Again, this shows
that we can induce first-order filling by tailoring (micro-patterning)
the substrate close to the wedge bottom. 
This may provide a practical means of reducing the fluctuation effects
which would otherwise dominate any continuous filling transition. 
This finding may have technological
implications for microfluidic devices. However, the borderline between 
first-order and critical filling depends on the specific range of the 
interactions. 
If the binding potential between the interface and the flat wall decays
faster than $1/z^4$, where $z$ is the local interfacial height above the
substrate, both regimes are separated by a tricritical point, as in the
case of contact binding potentials \cite{JM1,JM2}. On the other hand,
for longer-ranged binding potentials, a critical end point separates the 
first-order and critical filling transitions. We note that these two 
situations correspond exactly to the fluctuation-dominated 
and mean-field regimes for critical filling \cite{Parry,Parry4}. 

Our Paper is arranged as follows: In Section II we review briefly 
the phenomenology of wedge filling and introduce the breather mode 
interfacial model used in our study. The  definition of the path integral
used in our transfer matrix analysis is discussed in some detail. While
other formalisms have been forwarded they suffer from a number of
problems. As we shall show our definition is consistent with 
thermodynamic requirements (exact sum-rules), generalized random walk 
arguments and also the predictions of conformal invariance.
Section III is devoted to the analysis of wedge filling for contact 
binding potentials. Some of these results have been previously reported,
without derivation, in a brief communication \cite{JM1,JM2}. 
Section IV extends the transfer matrix analysis to the important practical
case of filling with long-ranged van der Waals forces. We conclude
with a brief discussion and summary. 

\section{The model}
\begin{figure}
\begin{center}
\includegraphics[width=9cm]{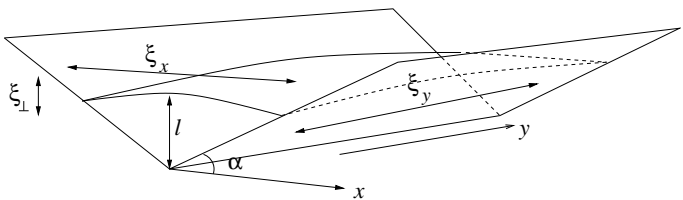}
\caption{Schematic illustration of a typical interfacial configuration in the
3D wedge geometry. The diverging lengthscales at the filling 
transition are highlighted.
\label{fig1}}
\end{center}
\end{figure}
Our starting point is the interfacial Hamiltonian pertinent to filling in
shallow wedges (small tilt angle $\alpha$) \cite{Rejmer}:
\begin{equation}
{\mathcal H}[l]=\int \int dx dy \left[\frac{\Sigma}{2}(\nabla z)^2+W(z-
\alpha|x|,x)\right]
\label{heff1}
\end{equation}
where $z(x,y)$ is the local height of the liquid-vapour interface relative
to the horizontal, $\Sigma$ is the liquid-vapour surface tension and
$W(z,x)$ is the binding potential between the liquid-vapour interface
and the substrate (see figure \ref{fig1}). Here after we assume the 
temperature defines the energy scale and set $k_B T=1$. To first approximation
we may suppose that $W(z,x)$ is independent of the position across the
wedge section, i.e.  $W(z,x)\approx W_\pi(z)$, where $W_\pi(z)$ is the 
binding potential due to a single flat substrate. Corrections may 
arise when the liquid adsorption is small enough however - a point we shall 
return to later. 

A mean-field analysis shows that locally the interface across the wedge is 
flat and that fluctuation effects are dominated by pseudo-one-dimensional 
local translations in the height of the filled region along the wedge 
(the \emph{breather modes}) \cite{Parry,Parry4}. Fluctuation effects at
filling can be studied by using an effective pseudo-one-dimensional wedge 
Hamiltonian which accounts only for the breather-mode excitations 
\cite{Parry,Parry4}:
\begin{equation}
{\mathcal H}_W[l]=\int d y\left\{\frac{\Lambda(l)}{2}
\left(\frac{dl}{dy}\right)^2
+V_W(l)\right\}
\label{heff2}
\end{equation}
where $l(y)=z(0,y)$ is the local height of the interface above the wedge
bottom. The effective bending term $\Lambda(l)$ resisting fluctuations 
along the wedge can be expressed as \cite{Parry,Parry4}:
\begin{equation}
\Lambda(l) \approx \frac{2\Sigma l}{\alpha}+\frac{2\tau}{\alpha^2}
\approx \frac{2\Sigma l}{\alpha}
\label{deflambda}
\end{equation}
where $\tau$ is the line tension associated with the contact lines between
the filled region and the substrate far from the wedge bottom. Note that  
for large $l$, we may neglect the $\tau$ contribution since 
$\Lambda(l)$ is proportional to the local interfacial height.
Similarly, for large $l$ the effective binding potential 
$V_W(l)$ takes the form 
\cite{Parry,Parry4}:
\begin{eqnarray}
V_W(l)\approx \frac{h(l^2-l_\pi^2)}{\alpha}+\frac{\Sigma(\theta^2-\alpha^2) 
(l-l_\pi)}{\alpha}\nonumber\\
+2\tau + \tau'+\int_{-(l-l_\pi)/\alpha}^{(l-l_\pi)/\alpha} dx \  
W_\pi(l-\alpha|x|)
\label{effbinding}
\end{eqnarray}
The first two terms corresponds to the bulk and surface thermodynamic
contributions required to form the filled liquid region \cite{Parry2}. 
Here $h$ denotes the bulk ordering field measuring deviations from
bulk two-phase coexistence, $\theta$ is the 
contact angle of the liquid drop at the planar wall-vapour interface and
$l_\pi$ is the equilibrium liquid layer thickness for a single planar wall. 
The line tension $\tau$ is defined as above, and $\tau'$ is the line tension
associated with the wedge bottom. Note that the line tension contributions 
are essentially independent of $l$ for $l \gg l_\pi$, so they become 
irrelevant in that limit. Finally, the last term corresponds to the binding 
potential contribution to $V_W$. Upon minimisation of $V_W(l)$, we recover 
the mean-field expression for the mid-point height (at bulk coexistence) 
\cite{Parry,Parry4}:
\begin{equation}
\frac{\Sigma \alpha^2}{2}=W_\pi(l)+\frac{\Sigma \theta^2}{2}\equiv 
\Delta W_\pi(l)
\label{MF}
\end{equation}
However, we stress that the form (\ref{effbinding}) for $V_W(l)$ 
is only valid for $l\gg l_\pi$. For $l\lesssim l_\pi$, both $\Lambda(l)$ and
$V_W(l)$ will behave in a different manner. Furthermore, we may control
the adsorption properties for small $l$ by micropatterning a stripe along
the wedge bottom, so as to weaken the local wall-fluid intermolecular 
potential. The interfacial binding potential is consequently strengthened, 
and under some conditions it may bind the liquid-vapour interface 
to the wedge bottom even at the filling transition boundary $\theta=\alpha$. 
Thus by introducing a line tension associated with the wedge bottom one
may induce first-order filling in the modified wedge provided the modification
is strong enough. As we shall see the lines of first-order and continuous 
filling transitions are separated by either a tricritical point or a 
critical end point depending on the range of the intermolecular forces. 

The quasi-one-dimensional character of the Hamiltonian means it 
is amenable to a transfer-matrix analysis. The 
partition function corresponding to this Hamiltonian
can be expressed as the following path integral \cite{Burkhardt}:
\begin{equation}
Z(l_b,l_a,Y)=\int {\mathcal D}l \exp(-{\mathcal H}_W[l]) 
\label{partfunc}
\end{equation}
However, the presence of a position-dependent stiffness coefficient makes the
definition of the partition function ambiguous. This problem was 
pointed out, but not satisfactorily resolved, in \cite{Bednorz} and is
intimately related to issues associated with the canonical quantization of
classical systems with a position-dependent mass \cite{Thomsen,Chetouani}.
In this paper we use the following definition of the partition function:
\begin{equation}
Z(l_b,l_a,Y)=\lim_{N\to \infty} \int dl_1\ldots dl_{N-1} \prod_{j=1}^N 
K(l_j,l_{j-1},Y/N)\label{partfunc2}
\end{equation}
where $l_0\equiv l_a$ and $l_N\equiv l_b$, and $K(l,l',y)$ is defined as:
\begin{eqnarray}
K(l,l',y)={\frac{\left(\Lambda(l)\Lambda(l')\right)^{1/4}}
{\sqrt{2 \pi y}}}
\textrm{e}^{-\frac{\sqrt{\Lambda(l)\Lambda(l')}}{2 y}(l-l')^2-
y V_W(l)}
\label{partfunc3}
\end{eqnarray}
The partition function $Z(l_b,l_a,Y)$ satisfies the differential 
equation
\begin{equation}
H_W Z(l_b,l_a,Y) = -\frac{\partial Z(l_b,l_a,Y)}{\partial Y}
\label{schrodinger1}
\end{equation}
with initial condition $Z(l_b,l_a,Y)\to \delta(l_b-l_a)$ as $Y\to 0$. 
The operator $H_W$ is defined as \cite{Chetouani}:
\begin{equation}
H_W \equiv -\frac{1}{2}\frac{\partial}
{\partial l_b} \left[\frac{1}{\Lambda(l_b)}\frac{\partial}
{\partial l_b}\right]
+ V_W(l_b)+\tilde{V}_W(l_b)
\label{defhw}
\end{equation}
where $\tilde{V}_W(l)$ is given by
\begin{equation}
\tilde{V}_W(l)=-\frac{1}{2\Lambda(l)}\left[\frac{3}{4}\left(\frac{\Lambda'(l)}
{\Lambda(l)}\right)^2-\frac{\Lambda''(l)}{2\Lambda(l)}\right]
\label{deftildev}
\end{equation}
and prime denotes differentiation with respect to argument. 
The solution of (\ref{schrodinger1}) can be 
expressed via the spectral expansion \cite{Burkhardt}:
\begin{equation}
Z(l_b,l_a,Y)=\sum_\alpha \psi_\alpha(l_b)\psi_\alpha^*(l_a)
\textrm{e}^{-E_\alpha Y}
\label{partfunc4}
\end{equation}
where $\{\psi_\alpha(l)\}$ is a complete orthonormal set of eigenfunctions
of the Hamiltonian operator $H_W$, with associated eigenvalues $E_\alpha$. 

Analogous to discussion of 2D wetting \cite{Burkhardt} we can now obtain 
the interfacial properties from the knowledge of the propagator $Z(l_b,l_a,Y)$.
In particular, the probability distribution function (PDF), $P_W(l)$, 
can be obtained as:
\begin{equation}
P_W(l,Y)=\lim_{L\to \infty} \frac{Z(l,l_{-L/2},Y+L/2)Z(l_{L/2},l,L/2-Y)}
{Z(l_{L/2},l_{-L/2},L)}
\label{defpw0}
\end{equation}
while the joint probability $P_W^{(2)}(l_1,l_2,Y_1,Y_2)$ 
of finding the interface at midpoint heights $l_1$ and $l_2$ at positions 
$Y_1$ and $Y_2(>Y_1)$, respectively, is:
\begin{eqnarray}
P_W^{(2)}(l_1,l_2,Y_1,Y_2)=
Z(l_2,l_1,Y_2-Y_1)\nonumber\\ \times
\lim_{L\to \infty} \frac{Z(l_1,l_{-L/2},Y_1+L/2)
Z(l_{L/2},l_2,L/2-Y_2)}
{Z(l_{L/2},l_{-L/2},L)}
\label{defpw2}
\end{eqnarray}
Further simplifications arises if we assume the existence of a bounded 
ground eigenstate $\psi_0$. Then, substitution of the spectral expansion
(\ref{partfunc4}) into (\ref{defpw0}) reads (for an infinitely long 
wedge) 
\begin{equation}
P_W(l)=|\psi_0(l)|^2 
\label{defpw}
\end{equation}
Similarly the excess wedge free-energy per unit length of an infinitely long
wedge is identified with the ground eigenvalue $E_0$.
The two-point correlation function $h(l_1,l_2,Y)$ 
can be obtained in a similar manner:
\begin{eqnarray}
h(l_1,l_2,Y)\equiv P_W^{(2)}(l_1,l_2,0,Y)-P_W(l_1)P_W(l_2)
\nonumber\\
=\sum_{\alpha\ne 0} \psi_\alpha^*(l_1)\psi_0(l_1) \psi_\alpha(l_2)\psi_0^*(l_2)
\exp\left[-(E_\alpha-E_0)Y\right]
\label{correlation}
\end{eqnarray}
For large separations this vanishes exponentially allowing us to determine the 
correlation length $\xi_y =(E_1-E_0)^{-1}$, where $E_1$ is 
eigenvalue corresponding to the first excited eigenstate. This result still
holds even if $E_1$ corresponds to the lower limit of the continuous part of 
the spectrum of $H_W$, although now the leading order of $h(l_1,l_2,Y)$ is 
not purely exponential but is modulated by a power of $Y$.

An interesting connection with 2D wetting problems can now be seen. 
Introducing the change of variables \cite{Yu}
\begin{equation}
\eta(l)=\int dl \sqrt{\Lambda(l)}\quad ; \quad  \phi(\eta)=\Lambda(l)^{-1/4} 
\psi\left(l(\eta)\right)
\label{changevar}
\end{equation}
the eigenvalue problem $H_W \psi_\alpha (l) = E_\alpha \psi_\alpha (l)$ 
transforms to the following Schr\"odinger-like equation:
\begin{equation}
-\frac{1}{2}\frac{d^2 \phi_\alpha(\eta)}{d\eta^2} + \left(V_W[l(\eta)]+
\tilde{V}_W^*(\eta)\right)\phi_\alpha(\eta)=E_\alpha \phi_\alpha(\eta)
\label{schrodinger2}
\end{equation}
where $\tilde V_W^*$ is defined as:
\begin{equation}
\tilde{V}_W^*(\eta)=-\frac{3}{32 \Lambda^2(l(\eta))}\left(
\frac{d \Lambda(l(\eta))}{d\eta}\right)^2+\frac{1}{8\Lambda(l(\eta))}
\frac{d^2\Lambda(l(\eta))}{d\eta^2}
\label{deftildev2}
\end{equation}
(\ref{schrodinger2}) shows that the filling problem can be mapped onto
a 2D wetting problem in the $\eta$ variable, under an effective binding 
potential. In our case $\Lambda(l)=2\Sigma l/\alpha$, so we obtain 
$\tilde V_W(l)=-3\alpha/16\Sigma l^3$. The change of variables  
(\ref{changevar}) leads to $\eta=\sqrt{8\Sigma/9\alpha} l^{3/2}$ and 
thus $\tilde V_W^*(\eta)=-5/72\eta^2$. Note that the use of the variable 
$\eta \propto l^{3/2}$ as the appropriate collective coordinate 
has been previously recognized in the literature 
\cite{Bednorz,Parry2}. However, our definition of the path integral
(\ref{partfunc2}) and (\ref{partfunc3}) leads to a novel term in the wedge 
binding potential, which will be essential in our study and ensures 
thermodynamic consistency. 

The filling potential $V_W(l)$  must fulfill some requirements. 
In order that the interface does not penetrate the substrate,  
$V_W(l)$ has a hard-wall repulsion for $l<0$. Consequently 
we need to impose an appropriate boundary condition on the 
eigenfunctions of $H_W$ at $l=0$.
The analytical expression of the boundary condition is obtained by a 
regularization procedure: we assume that the filling potential $V_W$ and 
the position-dependent stiffness $\Lambda$ are constant for $l<\xi_0$, 
where $\xi_0$ is some microscopic scale. Furthermore, we impose that 
$\Lambda(l)$ must be continuous at $l=\xi_0$, so $\Lambda(l<\xi_0)=2\Sigma 
\xi_0/\alpha$. On the other hand, the filling potential 
can take an arbitrary value $-U$. The latter square-well potential models the
modification of the filling potential $V_W(l)$ for small $l$ due to the
line tension associated with the wedge bottom. The eigenstates must fulfill the 
usual matching conditions that $\phi_\alpha$ and $\partial \phi_\alpha/
\partial l$ are continuous at $l=\xi_0$. Finally, we consider the 
appropriate scaling limit as $\xi_0\to 0$. 

The qualitative form of the filling potential depends on the order of the 
mean-field phase transition \cite{Parry,Parry4}. 
For critical filling with long-ranged intermolecular forces, 
the binding potential at bulk coexistence behaves as 
\begin{equation}
V_W(l)=\frac{\Sigma(\theta^2-\alpha^2)l}{\alpha}+\frac{2A}
{(p-1)\alpha}l^{1-p}+\ldots
\label{expansionvw}
\end{equation}
where $A$ is a Hamaker constant while the exponent $p$ depends on 
the range of the forces. Specifically, for non-retarded van der Waals forces 
$p=2$. For systems with short-ranged forces this is replaced by an exponential
decay $\sim \exp(-\kappa l)$ where $\kappa$ is an inverse bulk correlation
length. The presence of the fluctuation-induced filling 
potential $\tilde V_W(l) \propto l^{-3}$ in (\ref{schrodinger1})
gives rise to two distinct scenarios. For $p>4$ and large $l$, the direct 
contribution in (\ref{schrodinger1}) $\propto l^{1-p}$ is negligible
compared to the fluctuation-induced potential $\tilde V_W$ arising from the
position dependent stiffness. In this case we anticipate universal, 
fluctuation dominated behaviour. On the other hand, for $p<4$ and
large $l$ we can neglect $\tilde V_W$. Since the $l^{1-p}$ contribution to
the binding potential is now repulsive, we expect a qualitatively different
phase diagram. It is worthwhile noting that
both situations exactly correspond to the existence of a mean-field and 
a filling fluctuation-dominated regime for the filling transition 
predicted from heuristic arguments \cite{Parry,Parry4}.

\begin{figure}
\begin{center}
\includegraphics[width=9cm]{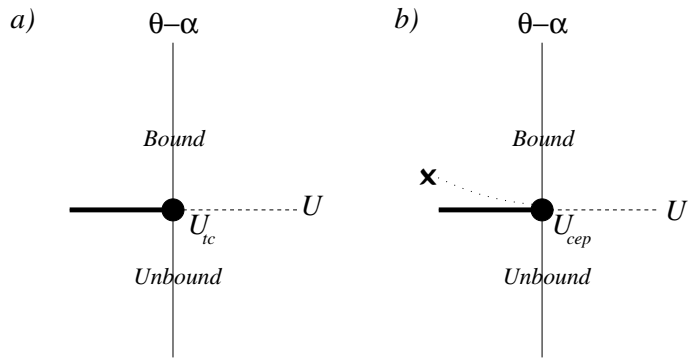}
\caption{Schematic phase diagrams for 3D wedge filling pertinent to: 
(a) short-range forces, (b) long-range forces. $U$ is the effective line 
tension strength, $\theta$ is the contact angle for the liquid on a planar 
substrate, and $\alpha$ the wedge angle. The filled circle locates the 
borderline between the first-order transition line (dashed line) and the 
second-order transition line (thick continuous line) between the bound and 
unbound states. These correspond to a tricritical point and a critical 
end-point for short-ranged and long-ranged forces, respectively. 
In the latter case, the first-order transition 
line continues to a first-order pseudo-transition line in the partial 
filling region (dotted line), which terminates at a pseudo-critical point 
(cross). See text for explanation. \label{fig2}}
\end{center}
\end{figure}

Fig. \ref{fig2} shows the schematic filling phase diagrams we expect 
for short-ranged (Fig. 2a) and long-ranged forces (Fig. 2b). Previous 
work for the contact potential case \cite{JM1,JM2} showed a similarity 
between 3D wedge filling (for $h=0$) and 2D wetting, with $\theta-\alpha$ 
playing the role of the ordering field and the effective line tension the role
of the wetting binding potential. The borderline between the first-order
and the second-order transition lines corresponds to a tricritical point
(analogous to the 2D critical wetting case). For long-ranged forces 
the analogy between 3D filling and 2D wetting still holds (see above). 
However, as the effective binding potential is repulsive for large $l$ at 
$\theta=\alpha$, the similarity must be established with 2D \emph{first-order} 
wetting. As for contact binding potentials, we expect the 
interface to be bound to the wedge bottom at $\theta=\alpha$ 
for a large well depth $U$ (corresponding to the line-tension contribution). 
As $U$ decreases, the interface will unbind along the $\theta=\alpha$ path. 
However, the interface must tunnel through a free-energy barrier to become 
unbound. Consequently, the borderline between the first-order and second-order
wedge filling is a critical end-point, where the spectator phase is the bound
state at $\theta=\alpha$. Actually, the connection with wetting phenomena also
leads naturally to this picture, since first-order wetting was previously
recognized as an interfacial critical-end point scenario \cite{Robledo}. 
In principle, we may expect the first-order 
transition line to continue in the partially filled region as the coexistence
between two bound states with different adsorptions. This thin-thick transition
is analogue to the prewetting line, and it should terminate at a critical 
point. However, the quasi-one-dimensional character of the wedge geometry
rules out this transition, as it is destroyed by breather-mode fluctuations. 
Nevertheless traces of this smeared transition may be found in the bimodal 
form of the interfacial height PDF (see later). 

We will consider two cases which correspond to different regimes of the
filling transition. The contact interaction will be studied as the paradigm
of the filling fluctuation-dominated regime. On the other hand, 
the van der Waals ($p=2$) case is analysed as a prototypical case of the 
mean-field regime. 

\section{Results for contact interactions}

\begin{figure}
\begin{center}
\includegraphics[width=9cm]{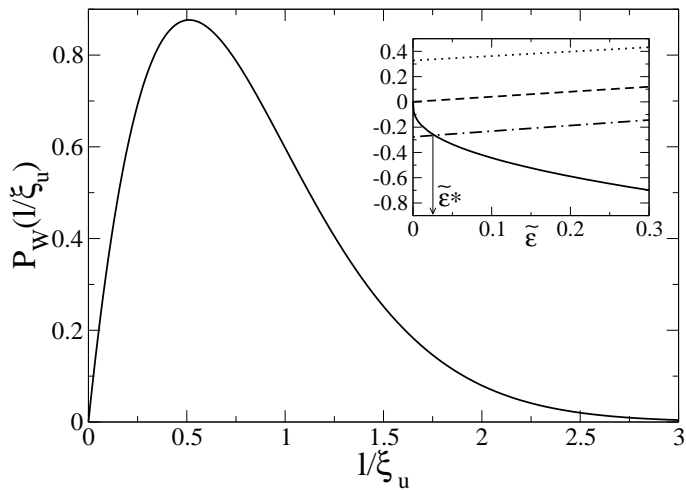}
\caption{Scaled probability distribution function $P_W(l/\xi_u)$ as a 
function of the scaled wedge midpoint interfacial height $l/\xi_u$ for 
$\theta=\alpha$. Inset: Plot of $\tilde \epsilon^{1/3}\textrm{Ai}'
(\tilde \epsilon^{1/3})/\textrm{Ai} (\tilde \epsilon^{1/3})$ (continuous line) 
and $\sqrt{u-\tilde \epsilon}\cot \sqrt{u-\tilde \epsilon}-1/2$ for 
$u=0.5<u_c$ (dotted line), $u=u_c=1.358$ (dashed line) and $u=2.0>u_c$ 
(dot-dashed line). For the latter case, the value of the reduced ground 
eigenvalue $\tilde \epsilon=\tilde \epsilon^*$ is highlighted. \label{fig3}}
\end{center}
\end{figure}

We consider first the case of short-ranged potentials. Some of our
results have been reported in a brief communication but without detailed
explanation \cite{JM1,JM2}. Here we provide full details of our transfer 
matrix solution and present new results for the form of the propagator.
At lengthscales much larger than that of the bulk correlation length $\xi_b$
we may write $V_W(l)=\Sigma(\theta^2-\alpha^2)l/\alpha$ for $l>0$ and allow for 
line tension arising from the wedge bottom via a suitable boundary 
condition at the origin. Analysis of (\ref{schrodinger1}) shows that
the short-distance behaviour of the eigenfunctions will be dominated by the 
$l^{-3}$ contribution to the effective filling potential. In fact, 
$\psi_\alpha(l)\sim l^{1/2}$ or $l^{3/2}$ as $l\to 0$, so the PDF 
function $P_W(l,\theta) \sim l$ or $l^3$ as $l\to 0$. We anticipate
that the latter behaviour corresponds to the critical filling transition 
as predicted by scaling arguments \cite{Parry2}, but the former will be
a completely new situation which, as we shall see, is related to the 
possibility of tricriticality. Turning to the critical filling transition 
first we note that the short-distance behaviour of the PDF, $P_W(l,\theta) \sim 
l^3$, which emerges from our analysis, is indeed the required,
thermodynamically consistent, result. This short-distance expansion ensures 
that the local density of matter near the wedge bottom contains a scaling
contribution that vanishes  $\propto T-T_f$, where $T_f$ is the filling 
temperature. This is the required singularity which emerges from an 
analysis of sum-rules connecting the local density near the wedge bottom to
(derivatives of) the excess wedge free-energy \cite{Parry3}.
This leads us to conclude that our definition of the path 
integral in (\ref{partfunc3}) is the correct one for  
the 3D wedge filling problem. As we shall see it also ensures that our 
model is conformally invariant.

\subsection{Wedge filling along the $\theta=\alpha$ path}
At the filling phase boundary ($\theta=\alpha$), the
problem can be mapped onto the intermediate fluctuation regime of 
2D wetting \cite{Lipowsky} via (\ref{schrodinger2}). 
We apply the regularization method described above, and consider the 
scaling limit $\xi_0\to 0$, $U\to \infty$ and $4\Sigma U \xi_0^3/\alpha\to u$. 
Under these conditions, the ground eigenvalue $E_0$ satisfies 
\begin{equation}
\sqrt{u-\tilde \epsilon}\cot \sqrt{u-\tilde \epsilon}-\frac{1}{2}=\tilde 
\epsilon^{1/3}\frac {\textrm{Ai}'\left(\tilde \epsilon^{1/3}\right)}
{\textrm{Ai}\left(\tilde \epsilon^{1/3}\right)}
\label{groundstate}
\end{equation}
where $\tilde \epsilon=-4\Sigma E_0 \xi_0^3/\alpha$ and $\textrm{Ai}(x)$ is the 
Airy function. Graphical solution of (\ref{groundstate}) (see inset of
figure \ref{fig3}) shows that there is a bound state with $E_0<0$ for $u>u_c$, 
where $u_c\approx 1.358$. Otherwise $E_0=0$ and there is no bound state. 
The existence of a bounded ground state at $\theta=\alpha$ implies that the 
filling transition is first-order for $u>u_c$, and critical for $u<u_c$. 
The explicit form of the PDF for $u>u_c$ in the thermodynamic limit is 
\begin{equation}
P_W(l,\theta=\alpha)=
\frac{6\sqrt{3}\pi}{\xi_u}\frac{l}{\xi_u}
\left[\textrm{Ai}\left(\frac{l}{\xi_u}\right)\right]^2
\label{PDF}
\end{equation}
where $\xi_u=\xi_0/\tilde \epsilon^{1/3}\propto \xi_0/(u-u_c)$ as $u\to u_c$.
Note that the lengthscale $\xi_u$ can be arbitrary large as $u\to u_c$.
Figure \ref{fig3} plots the PDF in terms of the scaling variable $l/\xi_u$. 
For small $l$, $P_W(l,\alpha)\sim l^{\gamma_u}$, with a short-distance 
exponent (SDE) $\gamma_u=1$. Asymptotically $P_W(l,\alpha)\sim 
\sqrt{l}\exp\left[-4(l/\xi_u)^{3/2}/3 \right]$ as $l\to \infty$. 

The mean interfacial mid-point height 
$l_W\equiv \langle l\rangle$ and roughness $\xi_\perp \equiv 
\sqrt{\langle l^2\rangle - \langle l\rangle^2}$ satisfy 
$l_W \sim \xi_\perp \sim \xi_u$ showing that, in the scaling limit, 
there is only one lengthscale controlling the fluctuations of the
interfacial height.
Finally, the correlation length along the wedge axis $\xi_y$ 
close to the filling transition can be obtained as $\xi_y=-1/E_0\equiv 4 \Sigma 
\xi_u^3/\alpha \propto (u-u_c)^{-3}$ since we can identify $E_1\equiv 0$.  

These observations indicate the emergence of a new relevant field 
(in the renormalization group sense) $t_u\propto (u-u_c)$, 
in addition to $t_\theta \propto\theta-\alpha$ and the bulk ordering field 
$h$. Thus the conditions $\theta=\alpha$, $u=u_c$ and $h=0$ correspond to a 
\emph{tricritical} point which separates the lines 
of first-order and critical filling transitions. The excess wedge 
free energy density $E_0$ vanishes as $E_0\sim t_u^{2-\alpha^u_W}
\sim t_u^3$ as $u$ tends to $u_c$ from above. Critical exponents 
for the divergence of the characteristic lengthscales can be defined as 
the tricritical point is approached along the $\theta=\alpha$ path:
\begin{equation}
l_w\sim t_u^{-\beta_W^u}\ ,\ \xi_\perp\sim t_u^{-\nu_\perp^u}\ ,\ 
\xi_y\sim t_u^{-\nu_y^u}
\label{critexp1}
\end{equation} 
Our results show that $\beta_W^u=\nu_\perp^u=1$ and $\nu_y^u=3$.
Finally, the effective wedge wandering exponent $\zeta_W=\nu_\perp^u/
\nu_y^u=1/3$, which coincides with its value for 
the critical filling transition \cite{Parry,Parry4}.

For $u\le u_c$ and $\theta=\alpha$, the interface is unbounded in the
thermodynamic limit. However, we can study the finite-size behaviour
of the droplet shape when the interface is pinned very close to the wedge
bottom at positions $y=\pm L/2$. Making use of results presented in the
Appendix, in particular (\ref{tricriticalz2}) and (\ref{criticalz2}), 
we find that the PDF at the tricritical ($u=u_c$) and critical ($u\ll u_c$) 
wedge filling transition are given by
\begin{eqnarray}
P_W^{u=u_c,L}(l,\theta=\alpha)=\frac{\lambda_L^2 l}{3^{1/3}\Gamma(2/3)}
\exp\left[-\frac{(\lambda_L l)^3}{9}\right]\label{droplet1}\\ 
P_W^{u\ll u_c,L}(l,\theta=\alpha)=\frac{\lambda_L^4 l^3}{3^{5/3}\Gamma(4/3)}
\exp\left[-\frac{(\lambda_L l)^3}{9}\right] 
\label{droplet2}
\end{eqnarray}
where 
\begin{equation}
\lambda_L=\left(\frac{16 \Sigma}{\alpha L}\right)^{1/3}
\left[1-\left(\frac{2Y}{L}\right)^2\right]^{-1/3}
\label{deflambdal}
\end{equation}
The typical droplet shape may be characterized by the most-probable position 
$l_{mp}(y)$ which follows from the relation 
$\partial P_W^L(l_mp)/\partial l=0$, or by the average shape $l_{av}(y)$
via the definition $l_{av}=\int_0^\infty dl l P_W^L(l)$ \cite{Burkhardt}. 
In all cases we find that the typical droplet shape follows 
$\lambda_L l=c$, where $c$ is a number which depends on the definition of the
typical shape and whether the pinning is at critical or tricritical filling. 
Consequently, the droplet shape must obey:
\begin{equation}
Y^2+\frac{4\Sigma L}{\alpha c^3}l^3=\frac{L^2}{4}
\label{dropletshape}
\end{equation}
Satisfyingly this shape is precisely that predicted by the requirement
of conformal invariance which can be used to map a droplet pinned
at just one end to one pinned at two points ($y=\pm L/2$)
\cite{Parry5}. This gives further indication that the definition of the 
measure used in our transfer matrix formulation is appropriate for the wedge 
geometry.

\subsection{The necklace model} 

A simple model can be introduced to understand the filling properties of
the wedge at $\theta=\alpha$. This model is a generalization of the
necklace model introduced for 2D wetting \cite{Fisher2,Fisher}. 
The interface is pinned to the wedge bottom along segments of
varying length but unbinds between them, forming liquid droplets.
We associate with each vertex a weight $v$ which is related to the
point tension between a bound and unbound state. 
Following the analysis described in \cite{Fisher2}, we introduce the 
generating function:
\begin{equation}
G(z)=\sum_{L=0}^\infty z^L Z_L
\label{defgenerating}
\end{equation}
where $Z_L$ is the interfacial canonical partition function of a segment of
length $L a$ (which we assume to be discretized in intervals of length $a$) 
and $z$ is an activity-like variable. The pure pinned ($A$) and unbound 
($B$) states have generating functions:    
\begin{eqnarray}
G_A=\sum_L Z_L^A z^L \sim \sum_L (s z)^L \label{defga}\\ 
G_B=\sum_L Z_L^B z^L \sim q_0 \sum_L \frac{(w z)^L}{L^\psi}
\label{defgb}
\end{eqnarray}
where $s=\exp(-u)$, with $u$ as an effective line tension in units
of $k_B T/a$. On the other hand, $w=\exp(-\tau_0)$, where $\tau_0$ is 
the reduced excess free energy per unit length of a long liquid droplet on 
the wedge, which we can assume to be zero by shifting the energy origin. 
Finally, $\psi$ is the exponent characterizing the first return of the 
interface to the wedge bottom.
This can be calculated from the $u\ll u_c$ limit corresponding to 
the completely filled regime. As shown in (\ref{scalingz}) and 
(\ref{criticalz3}) in the Appendix, $\psi=4/3$. Now the complete generating 
function is:
\begin{equation}
G(z)=G_A+G_AvG_BvG_A+\ldots=\frac{G_A(z)}{1-v^2G_A(z)G_B(z)}
\label{defg}
\end{equation}
Thus there will be a continuous phase transition at $u_c$ defined as 
\cite{Fisher2}:
\begin{equation}
u_c=-\ln\left[w(1-v^2G_c)\right]
\label{defuc}
\end{equation}
where $G_c=\sum_L Z_L^B/w^L$. Below $u_c$, the interface is completely unbound
while it remains pinned for $u>u_c$. The singularities of the various 
interfacial properties close to $u_c$ are characterized by critical 
exponents. In particular, the specific heat critical exponent $\alpha^u_W$ and
the longitudinal correlation length
critical exponent $\nu_y^u$ can be expressed in terms of the exponent
$\psi$ as \cite{Fisher2,Fisher} 
\begin{equation}
2-\alpha^u_W=\nu_y^u=\frac{1}{\psi-1}
\label{necklaceexponents}
\end{equation}
Substituting $\psi=4/3$ we find the correct values 
$2-\alpha^u_W=\nu_y^u=3$ derived earlier. 
In order to estimate the interfacial average height and roughness, we need
the interfacial height PDF in a liquid bubble, which is given by
(\ref{droplet2}). A similar argument to the one presented in  
\cite{Fisher2} leads to:
\begin{equation}
\xi_\perp\sim \frac{\overline{Y_B^{4/3}}}{\overline{Y_B}} \quad,\quad
l_w \sim \frac{\overline{Y_B^{7/3}}}{\overline{Y_B^2}} 
\label{necklaceexponents2}
\end{equation}
where $Y_B$ is the length of a liquid bubble, and $\overline{a}$ is the
average of $a$ in the pure liquid (B) phase. It is straightforward to see
that $l_w\sim \xi_\perp \sim \xi_y^{\zeta_W}$, with $\zeta_W=1/3$. 
Consequently, $\nu_\perp^u=\beta_W^u=1$, in agreement with our exact results. 
Note that in contrast to the 2D wetting case \cite{Fisher2,Fisher} the
exponent associated to the probability of first return  
$\psi\ne 2-\zeta_W$. This can be traced to the role played by the position 
dependent stiffness in the filling model (\ref{heff2})
which biases the random-walk-like motion of the interface. 
As pointed out previously \cite{JM2} it is remarkable
that the critical exponents for 3D critical and tricritical wedge filling
are identical to those anticipated for 2D complete and critical wetting
with random bond disorder. The necklace model provides an elegant means of
understanding this unusual dimensional reduction. 

\subsection{Wedge filling for $\theta>\alpha$}
\begin{figure}
\begin{center}
\includegraphics[width=9cm]{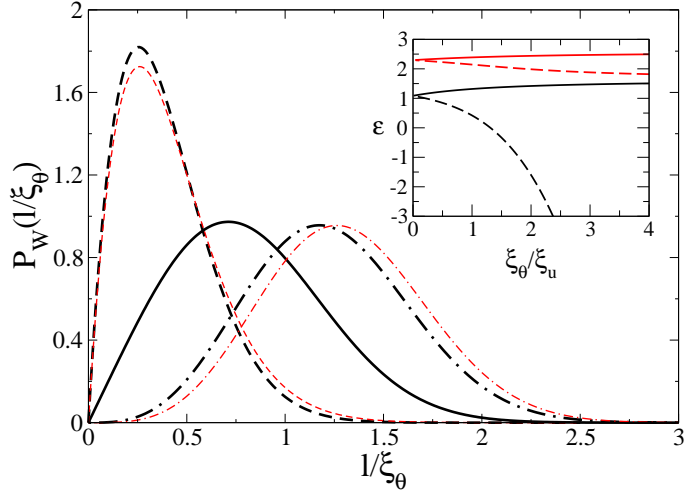}
\caption{Plot of the scaled PDF for $\epsilon_0=-1.5$ (thick dashed line),
$\epsilon_0\approx 1.086$, which corresponds to $u=u_c$ 
(thick continuous line) and $\epsilon_0\approx 1.639$, corresponding to
$u\ll u_c$ (thick dot-dashed line). For comparison, the PDF from  
(\ref{PDF}) with $\xi_u\approx 1.968\xi_\theta$ (which corresponds 
to $\epsilon_0=-1.5$, see inset) is also plotted (thin dashed line). 
Finally, the scaled PDF
obtained in \cite{Bednorz} is also shown (thin dot-dashed line).
Inset: Plot of the ground (black lines) and first-excited (red lines) 
reduced eigenvalues $\epsilon\equiv\Sigma E \xi_\theta^3/\alpha$ of $H_W$ as 
a function of $\xi_\theta/\xi_u$ for $u<u_c$ (continuous lines) and $u>u_c$ 
(dashed lines).
\label{fig4}}
\end{center}
\end{figure}
Now we extend some of our previous results to partial filling conditions, 
i.e. $\theta>\alpha$. The rescaling $\tilde \eta=(\Sigma/\alpha)^{1/4}
(\theta^2-\alpha^2)^{3/8}\eta/\sqrt{2}$, 
$\epsilon_\alpha=(\Sigma/\alpha)^{-1/2}(\theta^2-\alpha^2)^{-3/4}E_\alpha$
transforms the Schr\"odinger equation (\ref{schrodinger2}) 
into a parameter free form
\begin{equation}
-\frac{1}{4}\frac{d^2 \phi_\alpha(\tilde \eta)}{d\tilde \eta^2} + 
\left[\left(\frac{3\tilde \eta}{2} \right)^{2/3}-\frac{5}{144 \tilde 
\eta^2}\right]\phi_\alpha(\tilde \eta)=\epsilon_\alpha 
\phi_\alpha(\tilde \eta)
\label{schrodinger3}
\end{equation}
This leads directly to the critical behaviour of the mean mid-point 
height $l_W\equiv \langle l\rangle\sim (\theta-\alpha)^{-1/4}$ and the 
correlation length along the wedge $\xi_y\sim (\theta-\alpha)^{-3/4}$, 
in agreement with scaling arguments \cite{Parry,Parry4}, \emph{provided} that
the PDF is not singular, i.e. takes non-negligible values at finite values
of $\tilde \eta$. The PDF is obtained analogous to the case $\theta=\alpha$ 
case described above, from determination of the ground eigenstate $\psi_0$.
We find for the PDF 
\begin{eqnarray}
P_W(l,\theta>\alpha)
=C \frac{l}{\xi_\theta^2} \exp\left[
2\frac{l}{\xi_\theta}\left(\epsilon_0 - \frac{l}{\xi_\theta}
\right)\right]\nonumber\\
\left[H_{\frac{\epsilon_0^2}{4}-\frac{1}{2}}\left(\sqrt{2}\frac{l}{\xi_\theta}
-\frac{\epsilon_0}{\sqrt{2}}\right)\right]^2
\label{PDF2}
\end{eqnarray}
where $\xi_\theta\equiv\Sigma^{-1/2}[(\theta/\alpha)^2-1]^{-1/4}$, 
$\epsilon_0\equiv\Sigma E_0 \xi_\theta^3/\alpha$ is the reduced ground 
eigenvalue, $C$ is a normalization factor and $H_s(z)$ is the $s$-order Hermite
function \cite{Lebedev}. As $l\to 0$, the PDF vanishes (in general) 
like $P_W(l,\theta)\sim l$ while at large distances $P_W(l,
\theta)\sim l^{{\epsilon_0^2/2} -1}\exp[2l(\epsilon_0 - l/\xi_\theta)/ 
\xi_\theta]$ as $l\to \infty$. 

The value of the reduced ground eigenvalue $\epsilon_0$ depends on the 
boundary condition at $l=0$. In order to do this consistently we once
again turn to a regularization procedure: in the scaling limit 
($\xi_0\to 0$ and $4\Sigma U \xi_0^3/\alpha \to u$), the reduced 
eigenvalues are the solutions of the equation:
\begin{equation} 
\frac{\xi_\theta}{\xi_0}
\left(\sqrt{u}\ \textrm{cot}\sqrt{u}-\frac{1}{2}\right)=\epsilon + \left(\frac{ 
\epsilon^2}{\sqrt{2}}-\sqrt{2}\right)\frac {H_{\frac{\epsilon^2}{4}-
\frac{3}{2}}\left(-\frac{\epsilon}{\sqrt{2}}\right)}{H_{\frac{ 
\epsilon^2}{4}-\frac{1}{2}}\left(-\frac{\epsilon}{\sqrt{2}}\right)}
\label{groundstate2}
\end{equation}
where $\epsilon_0$ is the minimum value of them.
For $u$ close to $u_c$, we can expand the left-hand side of  
(\ref{groundstate2}) around $u_c$ as:
\begin{equation}
\frac{\xi_\theta}{\xi_0}
\left(\sqrt{u}\ \textrm{cot}\sqrt{u}-\frac{1}{2}\right)
\approx \pm \frac{\Gamma\left[-\frac{1}{3}\right]3^{-2/3}}
{\Gamma\left[\frac{1}{3}\right]}\frac{\xi_\theta}{\xi_u}
\label{expansion}
\end{equation}
where the positive (negative) sign corresponds to $u>u_c$ ($u<u_c$), 
respectively. This expression allows us to identify 
$\xi_u \propto \xi_0/|u-u_c|$ in
a manner completely consistent with our expression obtained for $u>u_c$ at 
$\theta=\alpha$. 

We obtain scaling behaviour for the wedge excess free-energy per unit length
$E_0$ (see also inset of figure \ref{fig4}): 
\begin{equation}
E_0=\frac{\alpha}{\Sigma\xi_\theta^3}
\epsilon_0^\pm\left(\frac{\xi_\theta}{\xi_u}\right)
\sim (\theta-\alpha)^{3/4} \epsilon_0^\pm 
\left[c\frac{u-u_c}{(\theta-\alpha)^{1/4}}
\right]
\label{scaling}
\end{equation}
where $c$ is an unimportant metric factor, and the sign corresponds to the
situations $u>u_c$ and $u<u_c$ as above. The scaling two functions 
$\epsilon_0^+$ and $\epsilon_0^-$ have the following properties:
\begin{eqnarray}
\epsilon_0^+(0)=\epsilon_0^-(0)=1.086\label{scalfunvalues1}\\
\epsilon_0^+(x\to +\infty)\sim -x^3\label{scalfunvalues2}\\
\epsilon_0^-(x\to +\infty)=1.639
\label{scalfunvalues3}
\end{eqnarray} 

The asymptotic behaviour of the PDF as the filling transition is approached,
i.e. $\theta \to \alpha$, is different for three situations: (i) $u>u_c$,
(ii) $u<u_c$ and (iii) $u=u_c$ (see figure \ref{fig4}). For the case (i), 
saddle-point asymptotic techniques \cite{Fyodorov} applied to the 
PDF (\ref{PDF2}) recover the expression for the PDF (\ref{PDF}). 
Consequently, the lengthscale governing the interfacial height and 
range of the breather-mode fluctuations is $\xi_u$, which remains
\emph{finite} as $\xi_\theta$ diverges. Thus on lengthscales compared to 
$\xi_\theta$ the PDF becomes a highly localized delta function located
at $l=0$. 
On the other hand, the first-excited eigenvalue scales as $E_1\propto 
\xi_\theta^{-3}$ (see inset of figure  \ref{fig4}), so the lateral correlation 
length also remains finite, $\xi_y\propto \xi_u^3$. 

For case (ii) we must take the limit (\ref{scalfunvalues3}) of 
the scaling
function $\epsilon_0^-$. Substitution into (\ref{PDF2}) leads to the 
asymptotic behaviour of the PDF as $\theta\to \alpha$. 
Now, the lengthscale which controls
both the mean interfacial height and roughness is $\xi_\theta$. It is
also interesting to note that $P_W(l)\sim l^3$ for $l\to 0$, so thermodynamic
consistency is assured. Our solution is different from the PDF reported in
\cite{Bednorz,Parry2} (see figure \ref{fig4}) although the global 
behaviour is qualitatively similar. On the other hand, the lateral 
correlation length $\xi_y$ has the asymptotic behaviour $\xi_y =(E_1-E_0)^{-1}
\sim \xi_\theta^3$.   

Finally, the PDF for case (iii) is obtained by substitution of the
limiting value (\ref{scalfunvalues1}) for $\epsilon_0$ into  
(\ref{PDF2}). Although the relevant lengthscales behave asymptotically as 
in the case (ii), the scaled PDF is different, as shown in figure \ref{fig4}. 
In particular, $P_W(l)\sim l$ as $l\to 0$. 

We can define the critical exponents for \emph{fixed} $u$ as:
\begin{equation}
E_0\sim t_\theta^{2-\alpha_W^\theta}\ ,\ 
l_w\sim t_\theta^{-\beta_W^\theta}\ ,\ \xi_\perp\sim 
t_\theta^{-\nu_\perp^\theta}\ ,\ \xi_y\sim t_\theta^{-\nu_y^\theta}
\label{critexp2}
\end{equation} 
where we define $t_\theta\propto \theta-\alpha$. For both
critical and tricritical filling, our analysis shows
that the critical exponents take the values $2-\alpha_W^\theta=3/4$,
$\beta_W^\theta=\nu_\perp^\theta=1/4$ and $\nu_y^\theta=3/4$, so
the wandering exponent is $\zeta_W=\nu_\perp^\theta/\nu_y^\theta=1/3$. 
These findings are in agreement with scaling predictions 
\cite{Parry,Parry4}.
Finally, we note that the tricritical gap exponent $\Delta^*$, which relates
the critical exponents along the $\theta=\alpha$ and fixed-$u$ paths, can
be obtained from the scaling form of $E_0$ (\ref{scaling}) as 
$\Delta^*=4$ \cite{JM2}.
\section{Results for dispersion forces}
\begin{figure}
\begin{center}
\includegraphics[width=9cm]{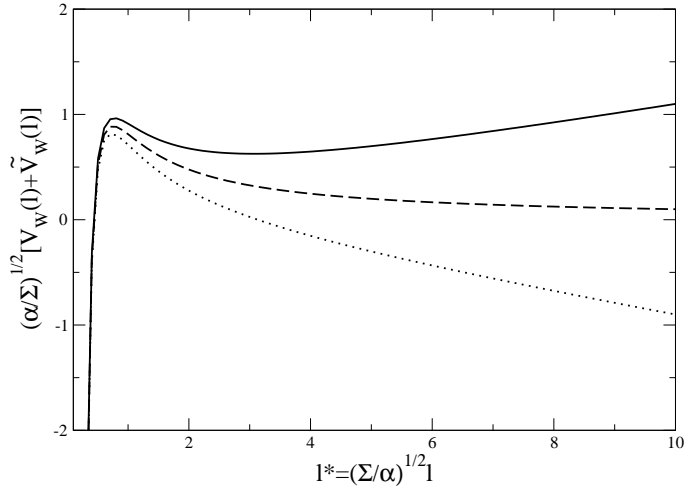}
\caption{The effective wedge binding potential $V_W(l)+\tilde V_W(l)$ 
for an effective Hamaker constant $2A/\alpha=1$ and: (i) $\theta^2-\alpha^2=0.1$
(continuous line), (ii) $\theta=\alpha$ (dashed line) and (iii) 
$\theta^2-\alpha^2=-0.1$ (dot-dashed line). 
\label{fig5}}
\end{center}
\end{figure}

Wedge filling in systems with short-ranged forces is representative of
the universality class of fluctuation dominated behaviour occurring if the
exponent in the binding potential $p>4$. In almost all practical
realizations of wedge filling however, long-ranged, van der Waals forces
from the fluid-fluid and solid-fluid intermolecular potentials will be
present. While exceptions to this may be found, for example in polymer
systems, the case of long-ranged forces is much more the rule than the
exception. We wish to understand how such long-ranged forces alter the
nature of the phase diagram and in particular the change from first-order
to continuous filling behaviour. We can allow for the presence of long-ranged 
forces through the binding potential (\ref{expansionvw}). 
For the three-dimensional case and non-retarded van der
Waals interactions the value of $p=2$. While higher order-terms are present
these will not play a significant role in determining the physics and can
be safely ignored. Here we show that the potential
\begin{equation} 
V_W(l)=\frac{\Sigma (\theta^2-\alpha^2)l}{\alpha} + \frac{2A}{\alpha l}
\label{vw-vdw}
\end{equation}
is amenable to exact analysis. The form of the total effective binding
potential $V_W(l)+\tilde V_W(l)$ is shown in figure \ref{fig5}. As one can 
see there is a local maximum for small $l$ which arises from the competition 
between the fluctuation-induced component and the van der Waals contribution.
A rough estimate for its location can be obtained by
setting $\theta=\alpha$, and a simple calculation leads to
\begin{equation}
l^*=\frac{3\alpha}{4\sqrt{2 \Sigma A}}
\label{maxl}
\end{equation}
For $\theta>\alpha$, a local minimum is obtained for larger $l$. This minimum
arises from the balance between the thermodynamic contribution 
$\Sigma(\theta^2-\alpha^2)l/\alpha$ and the van der Waals component 
of the effective binding potential. The position of the minimum is governed
for $\theta-\alpha\ll \alpha$ by the mean-field interfacial height:
\begin{equation}
l_W^{MF}=\sqrt{\frac{2A}{\Sigma(\theta^2-\alpha^2)}}
\label{lwmf}
\end{equation}
Since for long-ranged forces we anticipate that mean-field theory describes 
correctly the critical filling transition, a third lengthscale is the 
mean-field roughness, identified as \cite{Parry,Parry4}:
\begin{equation}
\xi_\perp^{MF}= \frac{\sqrt{\alpha}}{2}\left(\Sigma l_W^{MF} W'_\pi(l_W^{MF})
\right)^{-1/4}
\label{MFroughness}
\end{equation}
Since in our case $W_\pi(l)=-A/l^2$, we find, from (\ref{lwmf}) 
\begin{equation}
\xi_\perp^{MF}=\frac{\xi_\theta}{2}
\end{equation} 
where our definition of $\xi_\theta$ is unchanged from the previous Section. 
It is remarkable that the roughness is independent of the strength of 
van der Waals interactions.

The fluctuation-induced contribution to the wedge binding potential suggests
it may be possible to find a bound state at a lengthscale determined by $l^*$,
for $\theta\ge \alpha$. However no bound state is possible for $\theta<\alpha$:
any interface would tunnel through the barrier and unbind completely from 
the wedge bottom.   

Our analysis proceeds along the same lines as the previous Section. First we
investigate allowed states that exist at the filling phase boundary
$\theta=\alpha$, and then extend our analysis to the partial
filling regime $\theta>\alpha$.

\subsection{Wedge filling along the $\theta=\alpha$ path}

\begin{figure}
\begin{center}
\includegraphics[width=9cm]{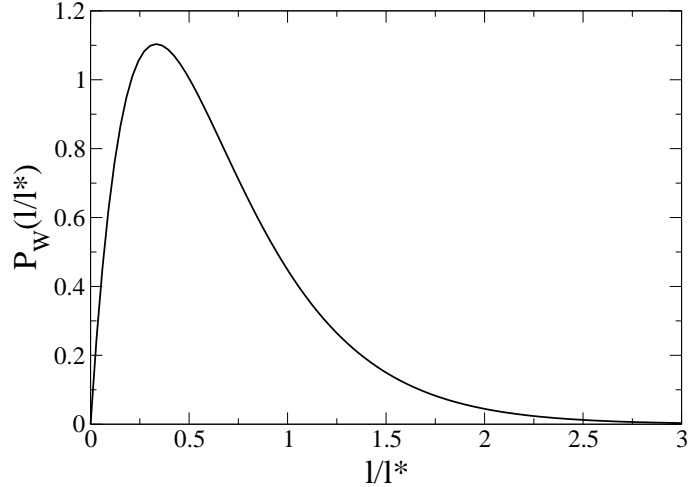}
\caption{Scaled probability distribution function $P_W(l/l^*)$  
for $\theta=\alpha$ and $u\to u_c$ in presence of dispersion forces. 
\label{fig6}}
\end{center}
\end{figure}

First we suppose $\theta=\alpha$ and search for a bound state $\psi_0(l)$ 
of $H_W$. As anticipated this is only possible 
if $E_0<0$, in which case the eigenfunction is given by 
\begin{equation}
\psi_0(l)\propto \sqrt{l} \textrm{Ai}\left(\frac{l}{\xi_u}+\frac{9\xi_u^2}
{4(l^*)^2}\right)
\label{groundstatevdw}
\end{equation}
where we define as previously $\xi_u=(-4\Sigma E_0/\alpha)^{-1/3}$.

In order to obtain $E_0$, we make use of the regularization procedure. In the 
scaling limit this satisfies the equation:
\begin{equation}
\sqrt{u-\tilde \epsilon}\cot \sqrt{u-\tilde \epsilon}-\frac{1}{2}=\tilde 
\epsilon^{1/3}\frac {\textrm{Ai}'\left(\tilde \epsilon^{1/3}
+\frac{9\xi_u^2}{4 (l^*)^2} \right)}
{\textrm{Ai}\left(\tilde \epsilon^{1/3}
+\frac{9\xi_u^2}{4 (l^*)^2} \right)}
\label{groundstatevdw2}
\end{equation}
where $\tilde \epsilon = (\xi_0/\xi_u)^3$. For values of $\tilde \epsilon$
not too close to zero, $\xi_u\sim \xi_0\ll l^*$ in the scaling limit, so
the second term in the Airy function argument can be neglected. Thus the 
solution for $\tilde \epsilon$ coincides with that corresponding to a 
contact binding potential. However, as $\tilde \epsilon\to 0$, a crossover
to a different situation is observed. In particular, for $\xi_u\gg l^*$ we
can make use of the asymptotic expansion of the Airy function for large
arguments: 
\begin{equation}
\textrm{Ai}(x)\sim \frac{\exp(-2x^{3/2}/3)}{2\sqrt{\pi}x^{1/4}}
\qquad x\to + \infty
\label{asymptairy}
\end{equation}
Substituting into (\ref{groundstatevdw2}), we find that
\begin{equation}
\sqrt{u-\tilde \epsilon}\cot \sqrt{u-\tilde \epsilon}-\frac{1}{2}\approx
-\frac{3\xi_0}{2l^*}-\frac{\tilde \epsilon l^*}{3\xi_0}\left(
\frac{l^*+3\xi_0}{3\xi_0}\right)
\label{groundstatevdw2-2}
\end{equation} 
Next we define $u_c$ as the value of $u$ at which $\tilde \epsilon=0$. It is 
straightforward to see that $u_c=1.358+{\cal O}(\xi_0/l^*)$, so in the scaling
limit the threshold for the existence of a bound
state is the same as for the contact binding potential.
Now we expand (\ref{groundstatevdw2-2}) for $u$ around $u_c$ and 
$\tilde \epsilon$ around zero. In the scaling limit, we obtain that:
\begin{equation}
\tilde \epsilon\sim \frac{\xi_0^2}{(l^*)^2}(u-u_c)
\label{dispersionrelation}
\end{equation}
Finally, the PDF in this regime can obtained from substitution
of the asymptotic relationship (\ref{asymptairy}) into  
(\ref{groundstatevdw}). After some algebra, the PDF reads: 
\begin{equation}
P_W(l)=\frac{9l}{(l^*)^2}\exp\left(-\frac{3l}{l^*}\right)
\label{PDFvdw}
\end{equation}
(see also figure \ref{fig6}). Since $l^*$ remains finite at $u_c$, 
the interface remains bound as $u$ approaches
$u_c$ from above, and suddenly unbinds for $u<u_c$. This identifies  
the point $h=0$, $\theta=\alpha$ and $u=u_c$ as a \emph{critical end point}
in the surface phase diagram. However, although $u$ is not a relevant 
field in the renormalization-group sense, the longitudinal correlation length 
$\xi_y=|E_0|^{-1}$ diverges as $(u-u_c)^{-1}$. A similar behaviour occurs 
within the subregime C of the intermediate fluctuation regime for 2D wetting
transitions \cite{Lipowsky}. 

\subsection{Wedge filling for $\theta>\alpha$}  
We now extend our results to $\theta>\alpha$ proceeding in the same way we
did for contact interactions. Again we can analytically obtain the ground
state $\psi_0(l)$ of $H_W$ as:
\begin{equation}
\psi_0(l)
\propto \sqrt{l} \exp\left[
\frac{l}{\xi_\theta}\left(\epsilon_0 - \frac{l}{\xi_\theta}
\right)\right]
H_{\frac{\epsilon_0^2}{4}-\frac{1}{2}-\frac{9\xi_\theta^2}{16(l^*)^2}}
\left(\sqrt{2}\frac{l}{\xi_\theta}
-\frac{\epsilon_0}{\sqrt{2}}\right)
\label{groundstatevdw3}
\end{equation}
where $\epsilon_0$, $\xi_\theta$ and $H_s(x)$ are defined as in
(\ref{PDF2}) for the contact binding potential. Note that the dependence
on the dispersion forces only appears at the order $s$ of the Hermite function
$H_s$. 

The regularization procedure leads to the following equation for the reduced
eigenvalues $\epsilon$ (for $u$ close to $u_c$):
\begin{eqnarray} 
\frac{\xi_\theta}{\xi_0}
\left(\sqrt{u}\ \textrm{cot}\sqrt{u}-\frac{1}{2}\right)
\approx \pm \frac{\Gamma\left[-\frac{1}{3}\right]3^{-2/3}}
{\Gamma\left[\frac{1}{3}\right]}\frac{\xi_\theta}{\xi_u}
-\frac{3\xi_\theta}{2l^*}\nonumber\\
= \epsilon + \left(\frac{ 
\epsilon^2}{\sqrt{2}}-\sqrt{2}-\frac{9\xi_\theta^2}{4\sqrt{2}(l^*)^2}
\right)\frac {H_{\frac{\epsilon^2}{4}-
\frac{3}{2}-\frac{9\xi_\theta^2}{16(l^*)^2}
}\left(-\frac{\epsilon}{\sqrt{2}}\right)}{H_{\frac{ 
\epsilon^2}{4}-\frac{1}{2}
-\frac{9\xi_\theta^2}{16(l^*)^2}}\left(-\frac{\epsilon}{\sqrt{2}}\right)}
\label{groundstatevdw4}
\end{eqnarray}
and $\epsilon_0$ is the minimum of the solutions. 
We can see that $\epsilon_0$ will depend now not only on the ratio 
$\xi_\theta/\xi_u$ and the sign of $u-u_c$, but also on $\xi_\theta/l^*$. 
Nevertheless as for the case with a contact binding potential, we are mainly
interested in the limit $\theta\to \alpha$, so that the 
lengthscale $\xi_\theta$ is very large. 

For $u>u_c$, saddle-point asymptotic techniques analogous to those
applied in the previous Section \cite{Fyodorov} show that the ground 
state of $H_W$ converges to the expression (\ref{groundstatevdw}) 
as $\theta\to \alpha$. Consequently we recover the results obtained earlier 
for the special case $\theta=\alpha$. This indicates that, for these
values of $u$, the wedge filling transition must be first-order.

For $u<u_c$ the non-existence of ground state for $\theta=\alpha$ indicates
that the filling transition is critical. We anticipate that mean-field 
theory will describe faithfully singularities at the critical wedge 
filling \cite{Parry,Parry4}. Thus we expect that the interfacial height 
PDF is centered around $l_W^{MF}$ with Gaussian fluctuations on the scale
of $\xi_\perp^{MF}$ representing the breather mode excitations.
The mean-field value of the excess free energy per unit length is given
by $V_W(l_W^{MF})$. Consequently, the mean-field value of the reduced
ground eigenvalue $\epsilon_0^{MF}$ is:
\begin{equation}
\epsilon_0^{MF}=\frac{3\xi_\theta}{2l^*}
\label{e0MF}
\end{equation}  
The shift of $\epsilon_0$ with respect to $\epsilon_0^{MF}$ due to 
breather-mode fluctuations $\Delta \epsilon_0$ can be estimated in the 
following way. We expand $V_W$ around its minimum up to quadratic order, 
so the shift may be estimated via:
\begin{equation}
\Delta E_0 \sim \frac{1}{2}V_W''(l_W^{MF})(\xi_\perp^{MF})^2
\label{shift}
\end{equation}  
implying that
\begin{equation}
\Delta \epsilon_0 \sim \frac{1}{\epsilon_0^{MF}}\sim \frac{l^*}{\xi_\theta}
\label{shift2}
\end{equation}
which vanishes as $\theta\to \alpha$. 

We can now proceed with a more formal derivation. Equation 
(\ref{groundstatevdw3}) can be written as:
\begin{equation}
\psi_0(l)\propto \sqrt{l} \exp(-x^2/2) H_s(x)
\label{groundstatevdw5}
\end{equation}
where we have defined:
\begin{eqnarray}
x&=&\frac{\sqrt{2}\Delta l}{\xi_\theta}-\frac{\Delta \epsilon_0}{\sqrt{2}}
\label{defx}\\
s&=&\frac{\epsilon_0^{MF} \Delta \epsilon_0-1}{2}+
\frac{(\Delta \epsilon_0)^2}{4}
\label{defs}
\end{eqnarray}
with $\Delta l\equiv l-l_W^{MF}$ and $\Delta \epsilon_0=\epsilon_0-
\epsilon_0^{MF}$. Note that (\ref{groundstatevdw5}) is the wavefunction
of the harmonic oscillator in the $x$ coordinate (in units of 
$\sqrt{\hbar/m\omega}$), modulated by the factor $\sqrt{l}$. 
As in the latter case, $\psi_0$ will increase exponentially as $x\to -\infty$,
i.e. $l\to 0$ and large $l_W^{MF}$, unless $s$ is a non-negative integer. 
This result is independent of the explicit value of $\xi_\theta/\xi_u$. 
Consequently, the shift $\Delta \epsilon$ for the lowest eigenvalues is
given by: 
\begin{equation}
\Delta \epsilon=\sqrt{(\epsilon_0^{MF})^2+4\left(n+\frac{1}{2}\right)}-
\epsilon_0^{MF}\approx \frac{2n+1}{\epsilon_0^{MF}}
\label{dispersionrelation2}
\end{equation} 
with $n$ a non-negative integer. The ground eigenstate will correspond to the
case $n=0$. The corresponding PDF becomes a Gaussian:
\begin{equation}
P_W(l)\approx \sqrt{\frac{2}{\pi\xi_\theta^2}}\exp\left(
-\frac{2(l-\langle l \rangle)^2}{\xi_\theta^2}\right)
\label{PDFvdw2}
\end{equation}
with $\langle l \rangle = l_W^{MF} + 2l^*/3 \approx l_W^{MF}$, and 
roughness $\xi_\perp=\xi_\theta/2$. Finally, the longitudinal correlation
length $\xi_y=(E_1-E_0)^{-1}\sim \xi_\theta^3 \epsilon_0^{MF}/2 \sim 
(\theta-\alpha)^{-1}$. 

Consequently, the explicit transfer matrix solution is in complete agreement
with the predicted mean-field values for the critical exponents when 
$p=2$ \cite{Parry,Parry4}: $2-\alpha_W^\theta=1/2$,
$\beta_W^\theta=1/2$, $\nu_y^\theta=1$ and $\nu_\perp^\theta=1/4$. 
 
\begin{figure}
\begin{center}
\includegraphics[width=9cm]{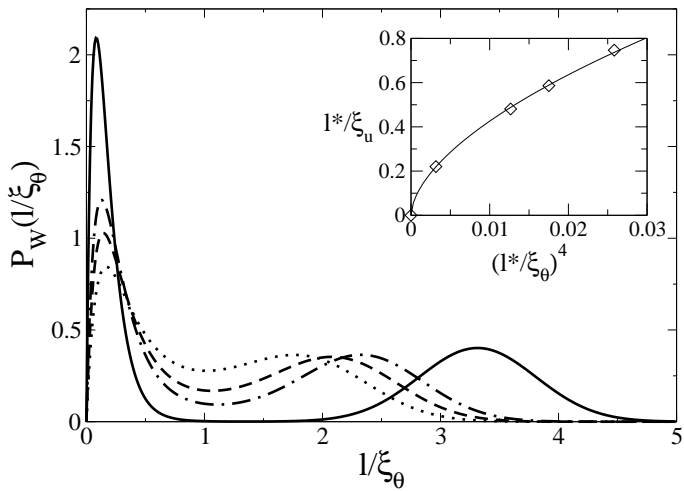}
\caption{Interfacial height PDF at the first-order pseudo-transition for 
$\xi_\theta/l^*=4.216$ (continuous line), $2.981$ (dot-dashed line),
$2.749$ (dashed line) and $2.494$ (dotted line). Inset: location of the
first-order pseudo-transition line. Note that $l^*/\xi_u\propto u_c-u$ 
(here $u<u_c$) and $(l^*/\xi_\theta)^4 \propto \theta-\alpha$. The diamonds
represent our calculated pseudo-coexistence values. The continuous line serves 
only as a guide for the eye.\label{fig7}}
\end{center}
\end{figure}

Finally, we search for the existence of the thin-thick transition line.
Our calculations show that there is no sharp phase transition for 
$\theta>\alpha$. However, we observe that the interfacial PDF becomes bimodal 
for $u<u_c$ and some range of values of $\theta>\alpha$. 
We can identify a first-order pseudo-transition line when the areas 
under each maximum of the PDF are equal (see Fig. \ref{fig7}). This line is the 
continuation to the partial filling region of the first-order filling 
transition line for $u<u_c$, and touches tangentially the filling transition
borderline at $u=u_c$. As $u$ decreases, the two maxima become closer, 
and eventually merge (the pseudo-critical point).  

\section{Discussion and conclusions}

In this paper we have reported analytical results for 3D wedge
filling transitions in the presence of short-ranged and long-ranged 
(van der Waals) interactions based on exact solution of the continuum 
transfer equations for a pseudo one-dimensional interfacial Hamiltonian. 
First-order and continuous (critical) filling are possible for both types 
of force depending on the strength of the line tension associated with 
the decoration of the wedge bottom. Our analytical
solution for the interfacial height PDFs at critical filling recovers the
known values of the critical exponents for critical filling for short and
long-ranged forces. In addition we have elucidated the nature of the
cross-over from first to critical filling which occurs at a specific value
of the wedge bottom line tension. This is qualitatively different for
systems with short and long ranged forces whose surface phase
diagrams are shown to have tricritical and critical end-points
respectively. To finish we argue that these results, which are
proto-typical of the fluctuation dominated and mean-field regimes
respectively, are qualitatively valid for any type of binding potential.
As already mentioned, the 3D wedge filling phenomenon can be mapped onto 
a 2D wetting problem with a new collective coordinate 
$\eta\propto l^{3/2}$. If we suppose that, at the filling phase boundary,
$l^3 V_W(l) \to 0$ for large $l$, we can make use of a renormalization-group 
arguments to determine the allowed behaviour. As $h$ and $t_\theta\propto 
\theta-\alpha$ are always relevant operators, we will restrict ourselves 
to the filling transition boundary $\theta=\alpha$, $h=0$. The effective 
2D wetting binding potential decays faster than $1/\eta^2$, so it becomes 
irrelevant in the renormalization-group sense \cite{Lipowsky2,Spohn}. The 
renormalization-group flows are dominated by the unstable fixed point and
the stable fixed point for the 2D wetting binding potential
$-5/72\eta^2$, which correspond to the tricritical and critical filling
transition, respectively. Thus, for large scales the only effect of such 
binding potentials that decay faster than $1/l^3$ is to renormalize the 
line tension associated with the wedge bottom.  

For long-ranged binding potentials i.e. those for which, at $\theta=\alpha$,
$l^3V_W(l)\to \infty$ for large $l$, we may resort to a mean-field analysis 
\cite{Parry,Parry4}. We expect that close to the filling 
transition the PDF is asymptotically a Gaussian characterized by the
mean-field interfacial height $l_W^{MF}$ and roughness $\xi_\perp^{MF}$.
However, if the next-to-leading order to the wedge binding potential
is determined by the short-distance repulsive part of $W_\pi(z)$, we find
a similar scenario to the one depicted in figure \ref{fig5}. In particular,
it is possible to bind the interface to the wedge bottom. The interfacial 
roughness will be controlled by the (microscopic) lengthscale $l^*\sim l_\pi$ 
corresponding to the maximum of the total effective binding potential. 
The existence of such a bound state, as well as the threshold to the critical 
filling transition will depend on the specific details of the interfacial 
binding potential. 

The predicted filling phenomenology presented in this paper can hopefully 
be checked experimentally or by computer simulations of more microscopic
models. Of course our predictions for contact (strictly short-ranged) forces
requires the elimination of van der Waals forces which are ubiquitous for
simple fluids. Here our results are most easily tested
using large scale Ising model simulations. In this case it should be
straightforward to induce first-order filling by weakening the local
spin-substrate interaction near the wedge bottom. In contrast our
predictions for first-order filling with van der Waals forces may well be
amenable to experimental verification sometime in the near future.
Taking an even broader perspective it may be that the chemical decoration of
a wedge bottom will provide a practical means of eliminating large scale
interfacial fluctuations. This may be of relevance to the construction of
microfluidic devices whose efficiency will depend crucially on the
control of fluctuation effects. 

\ack
J.M.R.-E. acknowledges partial financial support from Secretar\'{\i}a de
Estado de Educaci\'on y Universidades (Spain), co-financed by the
European Social Fund, and from the European Commission under Contract
MEIF-CT-2003-501042. A ``Ram\'on y Cajal'' Fellowship from the Spanish
Ministerio de Educaci\'on y Ciencia is also gratefully acknowledged. 
\appendix
\section{Evaluation of $Z(l_b,l_a;Y)$ at $\theta=\alpha$ for 
short-ranged forces.}

The evaluation of the interfacial properties at the filling boundary 
$\theta=\alpha$ is based on the knowledge of the partition function 
(\ref{partfunc}). In this Appendix we will evaluate explicitly the 
partition function $Z(l_b,l_a,Y)$ at the tricritical and critical points 
for short-ranged forces, and we will discuss some properties of the 
partition function for arbitrary $u$.

Our starting point is the solution of (\ref{schrodinger2}), for 
$\Lambda(l)=2\Sigma l/\alpha$ and $V_W(l)=0$, so the effective potential is 
$V_W^*(\eta)=-5/72\eta^2$. This Hamiltonian has $(1,1)$ deficiency 
indexes on the $(0,\infty)$ interval, so the solution is not unambiguosly 
defined, but instead depends on a parameter $c$ which defines the 
short-distance behaviour of the eigenfunctions, i.e.
\begin{equation}
\phi_\alpha(\eta)\sim c \eta^{5/6}+
\frac{2^{2/3}\Gamma(1/3)}{\Gamma(-1/3)}\eta^{1/6} 
\qquad \eta\to 0
\label{SDF}
\end{equation}
We define the partition function $\tilde Z(\eta_b,\eta_a,Y)$ via
the spectral expansion:
\begin{equation}
\tilde Z(\eta_b,\eta_a,Y)=\sum_\alpha \phi_\alpha(\eta_b)\phi^*_\alpha(\eta_a)
\textrm{e}^{-E_\alpha Y}
\label{defztilde}
\end{equation}
where the summation must be understood as an integral for the continuous 
part of the spectrum. From (\ref{changevar}), this partition function is 
related to $Z(l_b,l_a,Y)$ via:
\begin{equation}
Z(l_b,l_a,Y)=\sqrt{\frac{2\Sigma}{\alpha}} (l_b l_a)^{1/4}\tilde Z(\eta(l_b),
\eta(l_a),Y)
\label{connectionzztilde}
\end{equation}

The continuous part of the spectrum of any self-adjoint extension of our 
interfacial Hamiltonian corresponds to $E>0$. The corresponding scattering 
states can be expressed as \cite{Titchmarsh}:
\begin{equation}
\phi_E(\eta)=\frac{\sqrt{\eta}\left[c J_{1/3}(\sqrt{2E}\eta)-
(2E)^{1/3}J_{-1/3}(\sqrt{2E}\eta)
\right]}{\sqrt{c^2-c(2E)^{1/3}+(2E)^{2/3}}}
\label{scattering}
\end{equation}
where $J_s(x)$ is the s-order Bessel function of first kind. 

In addition, for $c>0$ there is a bounded eigenstate:
\begin{equation}
\phi_0(\eta)= \sqrt{\frac{3^{3/2} c^3\eta}{\pi}}K_{1/3}(c^{3/2}\eta) 
\label{boundstate}
\end{equation}
with associated eigenvalue $E_0=-c^3/2$, where $K_{s}(x)$ is the 
s-order modified Bessel function of second kind. 
Transforming back to the original variable $l=(9\alpha/8\Sigma)^{1/3} 
\eta^{2/3}$, the associated eigenfunction $\psi_0(l)$ is:
\begin{equation}
\psi_0(l)\propto \sqrt{\frac{l}{\xi_u}}\textrm{Ai}\left(\frac{l}{\xi_u}\right)
\label{boundstate2}
\end{equation}
where $\textrm{Ai}(x)$ is the Airy function, and $\xi_u$ is defined:
\begin{equation}
\xi_u=\left(\frac{\alpha}{2\Sigma}\right)^{1/3}\frac{1}{c}=
\left(-\frac{4\Sigma E_0}{\alpha}\right)^{-1/3}
\label{defc1}
\end{equation}
On the other hand, the longitudinal correlation length along the wedge
axis reads $\xi_y=2/c^3$. Comparison with the results obtained
by the regularization procedures in the text allows us to 
identify $c\propto (u-u_c)$.

For general $c$, we cannot perform the spectral integral. However, for $c=0$ 
($u=u_c$) and $c=-\infty$ ($u\ll u_c$) we have closed form expressions for 
$\tilde Z(\eta_b,\eta_a,Y)$ \cite{Erdelyi,Gradshteyn}:
\begin{eqnarray}
\tilde Z_{c=0}&=&\sqrt{\eta_a \eta_b} \int_0^\infty dE \textrm{e}^{-E Y}
J_{-\frac{1}{3}}(\sqrt{2E}\eta_a)J_{-\frac{1}{3}}(\sqrt{2E}\eta_b)\nonumber\\
&=&\frac{\sqrt{\eta_a \eta_b}}{Y}\exp\left(
-\frac{\eta_b^2+\eta_a^2}{2Y}\right)I_{-\frac{1}{3}}
\left(\frac{\eta_a\eta_b}{Y}\right)
\label{tricriticalz}\\
\tilde Z_{c=-\infty}&=&\sqrt{\eta_a \eta_b} \int_0^\infty dE 
\textrm{e}^{-E Y} J_{\frac{1}{3}}(\sqrt{2E}\eta_a)J_{\frac{1}{3}}
(\sqrt{2E}\eta_b)\nonumber\\
&=&\frac{\sqrt{\eta_a \eta_b}}{Y}\exp\left(
-\frac{\eta_b^2+\eta_a^2}{2Y}\right)I_{\frac{1}{3}}
\left(\frac{\eta_a\eta_b}{Y}\right)
\label{criticalz}
\end{eqnarray}
with the scaling property:
\begin{equation}
\tilde Z_{c=0,-\infty}=\frac{1}{\sqrt{Y}}
\tilde U_c\left(\frac{\eta_a}{\sqrt{Y}},
\frac{\eta_b}{\sqrt{Y}}\right)
\label{scalingtildez}
\end{equation}
Substituting into (\ref{connectionzztilde}) we obtain
\begin{eqnarray}
Z_{c=0} =\frac{l_a^* l_b^*}{3\xi_u Y^*}\exp\left(
-\frac{(l_a^*)^3+(l_b^*)^3}{9 Y^*}\right)
I_{-\frac{1}{3}}\left(\frac{2(l_a^* l_b^*)^{3/2}} {9 Y^*}\right)
\label{tricriticalz2}\\
Z_{c=-\infty} =\frac{l_a^* l_b^*}{3\xi_u Y^*}\exp\left(
-\frac{(l_a^*)^3+(l_b^*)^3}{9 Y^*}\right)
I_{\frac{1}{3}}\left(\frac{2(l_a^* l_b^*)^{3/2}} {9Y^*}\right)
\label{criticalz2}
\end{eqnarray}
where $l^*\equiv l/\xi_u$ and $Y^*\equiv Y/\xi_y$. Now the lengthscale
$\xi_u$ is arbitrary but $\xi_y=4\Sigma \xi_u^3/\alpha$.  
In both cases we have scaling such that 
\begin{equation}
Z_c=Y^{-1/3}U_c\left(\frac{l_a}{Y^{1/3}},
\frac{l_b}{Y^{1/3}}\right)
\label{scalingz}
\end{equation}
for $c=0$ and $-\infty$. For $Y\to \infty$, $U_c$ has the following asymptotic 
behaviour: 
\begin{eqnarray}
U_{c=0} &\propto&  
\sqrt{\frac{l_a}{Y^{1/3}}}\sqrt{\frac{l_b}{Y^{1/3}}}
\label{tricriticalz3}\\
U_{c=-\infty} &\propto&  
\left(\frac{l_a}{Y^{1/3}}\right)^{3/2}\left(\frac{l_b}{Y^{1/3}}\right)^{3/2}
\label{criticalz3}
\end{eqnarray}

To obtain results for arbitrary $c$, we will make use of the Krein formula 
\cite{Albeverio}, which relates the Green functions of 
different self-adjoint extensions of a closed symmetric operator. In our case, 
the Green function ${\cal Z}(l_b,l_a;E)$ is basically the Laplace transform 
of $Z(l_b,l_a;Y)$ with respect to $Y$:  
\begin{equation}
{\cal Z}(l_b,l_a;E)=\int_0^\infty dY \exp(EY)Z(l_b,l_a;Y)
\label{defgreen}
\end{equation}
For $c=0$ and $c=-\infty$ we have the closed expressions for the
Green function \cite{Erdelyi}:
\begin{eqnarray}
{\cal Z}_{c=0}&=&\frac{2 \xi_y l_a^* l_b^*}{3\xi_u} K_{\frac{1}{3}}
\left(\frac{2}{3}\sqrt{-E^*(l_>^*)^{3}}\right) 
I_{-\frac{1}{3}} \left(\frac{2}{3}\sqrt{-E^*(l_<^*)^3}\right)
\label{green1}\\
{\cal Z}_{c=-\infty}&=&\frac{2 \xi_y l_a^* l_b^*}{3\xi_u} K_{\frac{1}{3}}
\left(\frac{2}{3}\sqrt{-E^*(l_>^*)^{3}}\right) 
I_{\frac{1}{3}} \left(\frac{2}{3}\sqrt{-E^*(l_<^*)^{3}}\right)
\label{green2}
\end{eqnarray}
where $l_>^*$ and $l_<^*$ are the largest and smallest between $l_a^*$ and 
$l_b^*$, respectively, and $E^*\equiv E\xi_y$. 

To continue, we \emph{define} the lengthscales $\xi_u$ and $\xi_y$ 
for each $c$ as:
\begin{eqnarray}
\xi_u&=&\left(\frac{\alpha}{2\Sigma}\right)^{\frac{1}{3}}\frac{1}{|c|} 
\label{defc2}\\
\xi_y&=&\frac{2}{|c|^3}
\end{eqnarray}
Note that $\xi_u$ and $\xi_y$ reduce to the relevant correlation lengthscales
for $c>0$.
Application of the Krein formula leads to the following expression for the
Green function corresponding to an arbitrary $c$:
\begin{eqnarray}
{\cal Z}_c(l_b,l_a,E)={\cal Z}_{-\infty}(l_b,l_a,E)\nonumber\\
+g l_a^* l_b^* K_{\frac{1}{3}}\left(\frac{2}{3}\sqrt{-E^*(l_a^*)^{3}}\right)
K_{\frac{1}{3}}\left(\frac{2}{3}\sqrt{-E^*(l_b^*)^{3}}\right)
\label{kreinformula}
\end{eqnarray}
where $g\equiv g(E,\Sigma,\alpha)$ is obtained by imposing that the 
short-distance behaviour of $\cal Z$ is consistent with the boundary 
condition  
(\ref{SDF}). After some algebra, we obtain the following expression:
\begin{eqnarray}
{\cal Z}_c=\frac{{\cal Z}_{-\infty}}{1\mp(- E^*)^{1/3}}
\mp\frac{(- E^*)^{1/3}{\cal Z}_0}
{1\mp (- E^*)^{1/3}}
\label{kreinformula2}
\end{eqnarray}
where the negative sign corresponds to $c>0$, and the positive sign to $c<0$.
It is worthwhile to note that the dependence on the boundary condition, i.e.
$c$, has been absorbed into the lengthscales $\xi_u$ and $\xi_y$.
We can formally invert (\ref{kreinformula2}):
\begin{equation}
Z_c(l_b,l_a,Y)=\frac{1}{2\pi i} \int_{\gamma-i\infty}^{\gamma+i\infty}
dE \exp(-E Y) {\cal Z}_c(l_b,l_a,E)
\label{inverselaplace}
\end{equation}
with $\gamma<\min(0,-c^3/2)$. Note that there is a branch point of 
${\cal Z}_c$ at $E=0$. In addition, there is a single pole for $E=-c^3/2$ 
for $c>0$.
The scaling properties of ${\cal Z}_{c}$ (\ref{kreinformula2}) 
lead to the following scaling:
\begin{equation}
Z_c(l_b,l_a,Y)=\frac{1}{\xi_u}
Z_\pm \left(\frac{l_b}{\xi_u},\frac{l_a}{\xi_u},\frac{Y}{\xi_y}
\right)
\label{scalingzc}
\end{equation}

(\ref{inverselaplace}) provides a means of obtaining the asymptotic 
behaviour of $Z_c$ as $Y\to \infty$. For $c>0$, the dominant
contribution comes from the pole of ${\cal Z}_c$ at $E=-c^3/2$, so:
\begin{equation}
Z_c(l_b,l_a,Y)\sim \frac{6\sqrt{3}\pi}{\xi_u}\sqrt{l_a^* l_b^*}
\textrm{Ai}(l_a^*)\textrm{Ai}(l_b^*)\exp(Y^*)
\label{asympt1}
\end{equation}
in complete agreement with our previous analysis. Thus $\xi_u$ and $\xi_y$
are true correlation lengths.

For $c<0$, the branch point of ${\cal Z}_c$ at $E=0$ controls the large-$Y$ 
behaviour of $Z_c$. We expand ${\cal Z}_c$ around $E=0$: 
\begin{eqnarray}
{\cal Z}_c&\sim& \frac{\xi_y l_a^* l_b^*}{\xi_u} \Bigg[\left( 
\sqrt{\frac{l_<^*}{l_>^*}}-\frac{3^{2/3}\Gamma\left(\frac{1}{3}\right)}
{\Gamma\left(-\frac{1}{3}\right)\sqrt{l_a^* l_b^*}}\right)\nonumber\\
&+&(-E)^{1/3}
\left(\frac{\Gamma\left(-\frac{1}{3}\right)
\sqrt{l_a^* l_b^*}}{3^{2/3}\Gamma\left(\frac{1}{3}
\right)}-\sqrt{\frac{l_<^*}{l_>^*}}-\sqrt{\frac{l_>^*}{l_<^*}}
+\frac{3^{2/3}\Gamma\left(\frac{1}{3}\right)}
{\Gamma\left(-\frac{1}{3}\right)\sqrt{l_a^* l_b^*}}\right)
\nonumber\\&+&
{\cal O}\left[(-E)^{2/3}\right]\Bigg]
\label{asympt2}
\end{eqnarray}  
Thus $Z_c$ has for large $Y$ the following asymptotic behaviour:
\begin{eqnarray}
Z_c &\sim& 
-\frac{l_a^* l_b^*}{\xi_u \Gamma\left(-\frac{1}{3}\right)
(Y^*)^{4/3}} \Bigg[
\frac{\Gamma\left(-\frac{1}{3}\right)
\sqrt{l_a^* l_b^*}}{3^{2/3}\Gamma\left(\frac{1}{3}
\right)}-\sqrt{\frac{l_<^*}{l_>^*}}-\sqrt{\frac{l_>^*}{l_<^*}}
\nonumber\\
&+&\frac{3^{2/3}\Gamma\left(\frac{1}{3}\right)}
{\Gamma\left(-\frac{1}{3}\right)\sqrt{l_a^* l_b^*}}\Bigg]
+\ldots
\label{asympt3}
\end{eqnarray} 
Consequently, we find that for $l_{a}^*,l_{b}^*\gg 1$ but $\sqrt{l_a^* l_b^*}/
(Y^*)^{1/3}$ small, the asymptotic behaviour of $Z_c$ is the same as for 
the $c=-\infty$ case, i.e. critical filling. For the $c<0$ case $\xi_u$ 
and $\xi_y$ are not correlation lengths, but the (microscopic) lengthscales 
for $l$ and $Y$, respectively, so the critical filling behaviour is observed 
over larger scales than these. 

In order to obtain the asymptotics of the correlation function 
$h(l_b,l_a,Y)$ for $c>0$, we also need the next-to-leading order dependence 
of $Y$ that emerges from $Z_c(l_b,l_a,Y)$. Our approach provides a systematic 
means of obtaining such correction which can be obtained in a similar manner 
as before. If we denote by $Z_c^l$ the leading contribution to $Z_c$ given by
(\ref{asympt1}), the first correction is given by: 
\begin{eqnarray}
Z_c-Z_c^l&\sim&
-\frac{l_a^* l_b^*}{\xi_u \Gamma\left(-\frac{1}{3}\right)
(Y^*)^{4/3}} \Bigg[
\frac{\Gamma\left(-\frac{1}{3}\right)
\sqrt{l_a^* l_b^*}}{3^{2/3}\Gamma\left(\frac{1}{3}
\right)}+\sqrt{\frac{l_<^*}{l_>^*}}+\sqrt{\frac{l_>^*}{l_<^*}}
\nonumber\\
&+&\frac{3^{2/3}\Gamma\left(\frac{1}{3}\right)}
{\Gamma\left(-\frac{1}{3}\right)\sqrt{l_a^* l_b^*}}\Bigg]
+\ldots
\label{nexttoleading}
\end{eqnarray}

\end{document}